\documentclass[]{sig-alternate}\pdfoutput=1

\usepackage[utf8]{inputenc} 
\usepackage[table,xcdraw]{xcolor}
\usepackage{graphicx} 

\usepackage{balance}
\usepackage{paralist}

\makeatletter
\def\@copyrightspace{\relax}
\makeatother

\usepackage[]{pdfcomment}
\newcommand{\TODO}[1]{\textcolor{red}{#1}\pdfcomment[color=yellow,open=false]{#1}}\newcommand\todo\TODO

\newcommand{\mycode}[1]{{\small \texttt{#1}}\xspace}

\newcommand{\jgenprog}{jGenProg\xspace}
\newcommand{\jkali}{jKali\xspace}

\newcommand{\nbAllBugs}{224\xspace}

\newcommand{\nbFixedBugs}{47\xspace} 
\newcommand{\nbFixedBugsByKali}{22\xspace}
\newcommand{\nbFixedBugsByNopol}{35\xspace}
\newcommand{\nbFixedBugsByGenprog}{27\xspace}
\newcommand{\nbFixedBugsByAll}{12\xspace}
\newcommand{\nbDifferentPatches}{84\xspace}
\newcommand{\nbAnalyzedPatchCorrect}{11\xspace}
\newcommand{\nbAnalyzedBugCorrect}{9\xspace}
\newcommand{\nbAnalyzedGenprogCorrect}{5\xspace}
\newcommand{\nbAnalyzedIncorrect}{61\xspace}
\newcommand{\nbAnalyzedUnknown}{12\xspace}
\newcommand{\nbAnalyzedReaEasy}{61\xspace}
\newcommand{\nbAnalyzedDiffHardExpert}{21\xspace}
\newcommand{\experimenttimeindays}{17.6\xspace}
\newcommand{\averageExecutionTimeWithPatch}{14.8\xspace}

\usepackage{comment}

\usepackage[]{hyperref}
\usepackage{xspace}
\def\ourif{{\sc if}\xspace}

\usepackage{listings}
\usepackage{multirow}
\usepackage{amsmath}

\title{
Automatic Repair of Real Bugs: An Experience Report on the Defects4J Dataset 
}

\conferenceinfo{ICSE}{'16}
\CopyrightYear{2016}

\numberofauthors{1}
\author{
Matias~Martinez$^{1,2}$
,
Thomas~Durieux$^{1,2}$
,
Jifeng~Xuan$^3$
,
Romain Sommerard$^1$
,
Martin~Monperrus$^{1,2}$
\\
\\ (1) University of Lille, (2) INRIA, (3)  Wuhan University
}

\begin{document}
\maketitle

\begin{abstract} 
Defects4J is a large, peer-reviewed, structured dataset of real-world Java bugs. Each bug in Defects4J is provided with a test suite and at least one failing test case that triggers the bug.
In this paper, we report on an experiment to explore the effectiveness of automatic repair on Defects4J.
The result of our experiment shows that \nbFixedBugs bugs of the Defects4J dataset can be automatically repaired by state-of-the-art repair. 
This sets a baseline for future research on automatic repair for Java.
We have manually analyzed \nbDifferentPatches different patches to assess their real correctness.
In total, \nbAnalyzedBugCorrect real Java bugs can be correctly fixed with test-suite based repair.
This analysis shows that test-suite based repair suffers from under-specified bugs, for which trivial and incorrect patches still pass the test suite.
With respect to practical applicability, it takes in average \averageExecutionTimeWithPatch minutes to find a patch.
The experiment was done on a scientific grid, totaling \experimenttimeindays days of computation time. All their systems and experimental results are publicly available on Github in order to facilitate future research on automatic repair.

\end{abstract}

\section{Introduction}

Automatic software repair is the process of automatically fixing bugs.
Test-suite based repair, notably introduced by GenProg \cite{le2012genprog}, consists in synthesizing a patch that passes a given test suite with at least one failing test case. 
In this recent research field, few empirical evaluations have been made to evaluate the practical ability of current techniques to repair real bugs.
For instance, Le~Goues et al. \cite{le2012systematic} reported on an experiment where they ran the GenProg repair system on 105 bugs in C code. 

The key for a valuable empirical evaluation of automatic repair is a good dataset of bugs. 
Here, ``good'' means that the bugs are real (as opposed to seeded) and in large software applications (as opposed to small programs).
In the context of test-suite based repair, the bugs must also come with a test suite that encodes the expected behavior.
Defects4J is such a dataset \cite{JustJE2014}, which consists of 357 real-world Java bugs. It has been peer-reviewed, is publicly available, and is structured in a way that eases systematic experiments. Each bug in Defects4J comes with a test suite including failing test cases.
\emph{To explore whether automatic repair can be applied in practice, this paper asks the following question: could bugs in Defects4J be repaired with state-of-the-art repair approaches?}

But actually, a concrete bug cannot be repaired by a ``repair approach''. It is repaired by a ``repair tool''. For instance, the same term ``GenProg'' refers to both the approach and the tool, while they are different. To repair the bugs of Defects4J, we need executable tools. 

However, leaving aside the repair system developed in our group \cite{demarco2014automatic}, there are no other available test-suite based repair tools for Java, outside Arcuri's pioneering prototype \cite{DBLP:conf/cec/ArcuriY08} which is a small project incompatible with the scale and complexity of Defects4J.
So we chose to re-implement two key repair approaches. Re-implementing a repair system that works on real test suites and real code is a significant engineering effort. We have re-implemented GenProg \cite{le2012genprog} and Kali \cite{qi2015efficient}.
The motivation for choosing those two is as follows. First, GenProg is arguably a baseline in the field.
Second, Kali will help us to assess the quality of the test suites in Defects4J. It is a repair system only based on code deletion \cite{qi2015efficient} whose main goal is to identify under-specified bugs.
An ``under-specified'' bug is a bug for which the test cases that specify the expected behavior are weak. These test cases have a low coverage and bad assertions.
Indeed, if Kali fixes a failing test case by removing some code, it often means that the buggy code contained unspecified functionality. 
In the rest of this paper, we will use \jgenprog and \jkali to refer to these repair tools, in addition to 
Nopol \cite{demarco2014automatic,nopoljournal}, our repair tool based on speculative execution and code synthesis.
All of them are publicly available on Github.

In this paper, we present the results of an evaluation experiment consisting of running \jgenprog, \jkali, and Nopol on the bugs of Defects4J. 
Our experiment aims to answer to the following Research Questions (RQs):

\textbf{RQ1}. \textit{Can the bugs of the Defects4J dataset be fixed with the considered repair techniques?}
Answering this question is essential to consolidate the field of automatic repair. First, previous evaluations of automatic repair techniques were made on a bug dataset that was specifically built for the evaluation of those techniques. In other words, the authors of a technique and the authors of its evaluation dataset were the same. This increases the risk of potential biases due to the cherry-picking data. On the contrary, we are not authors of the Defects4J dataset. Second, while previous work has shown that real bugs in large scale C code\cite{le2012systematic,qi2015efficient,Long15} can be repaired, there is no reproducible work showing that real bugs from large scale Java projects can be repaired.

\textbf{RQ2}. \textit{In test-suite based repair, are the generated patches correct, beyond passing the test suite?}
By ``correct'', we mean that the patch is meaningful, really fixes the bug, and is not a partial fix that only works for the input data encoded in the test cases.
Indeed, a key concern behind test-suite based repair is whether test suites are acceptable to drive the generation of correct patches, where correct me acceptable. Since the inception of the field, this question has been raised many times and is still a hot question: Qi et al.'s recent results \cite{qi2015efficient} show that most GenProg's patches on the classical GenProg benchmark of 105 bugs are incorrect. We will answer RQ2 with a manual analysis of patches synthesized for Defects4J. 

\textbf{RQ3}. \textit{Which bugs in Defects4j are under-specified?}
For those bugs, current repair approaches fail to synthesize a correct patch due to the lack of test cases. Those bugs are the most challenging bugs: to automatically repair them, one needs to reason on the expected functionality below what is encoded in the test suite, to take into account a source of information other than the test suite execution.

\textbf{RQ4}. \textit{How long is the execution time of each repair approach?}
The answer to this question also contributes to assess
the practical applicability of automatic repair on real code.

Our experiment considers \nbAllBugs bugs that are spread over 231K lines of code and 12K test cases in total. We ran the experiment for over \experimenttimeindays days of computational time on Grid'5000 \cite{grid5000}, a large-scale grid for scientific experiments. 

Our contributions are as follows:

\begin{itemize}

\item \textbf{Answer to RQ1}. The Defects4J dataset contains bugs that can be automatically repaired with state-of-the-art techniques. Our implementations of \jgenprog, \jkali, and Nopol fix together \nbFixedBugs out of \nbAllBugs bugs with \nbDifferentPatches different patches. Nopol is the technique that fixes the largest number of bugs (\nbFixedBugsByNopol/\nbFixedBugs); some bugs are repaired by all three considered repair approaches (\nbFixedBugsByAll/\nbFixedBugs). This work can be viewed as a baseline for future usage of Defects4J in automatic repair research.

\item \textbf{Answer to RQ2}. Our manual analysis of all \nbDifferentPatches generated patches shows that \nbAnalyzedPatchCorrect/\nbDifferentPatches are correct, \nbAnalyzedIncorrect/\nbDifferentPatches are incorrect, and \nbAnalyzedUnknown/\nbDifferentPatches require a domain expertise, which we do not have. The incorrect patches tend to overfit the test cases. This is a novel piece of evidence that either the current test suites are too weak or the current automatic repair techniques are too dumb. 

\item \textbf{Answer to RQ3}. Defects4J contains very weakly specified bugs. Correctly fixing those bugs by an automatic repair approach that reasons beyond the test suite execution, using other sources of information, can be considered as the next milestone for the field.

\item \textbf{Answers to RQ4}. The process of searching for a patch is a matter of minutes for a single bug (RQ4). This is an encouraging piece of evidence for this research have an impact on practitioners.

\end{itemize}

For sake of open science and reproducible research, our code and experimental data are publicly available on Github:\\ 
\url{http://github.com/Spirals-Team/defects4j-repair/},\\
\url{http://github.com/SpoonLabs/nopol},\\ \url{http://github.com/SpoonLabs/astor}

The remainder of this paper is organized as follows. Section \ref{sect:background} provides the background of test-suite based repair and the dataset. Section \ref{sect:protocol} presents our experimental protocol. Section \ref{sect:result} details answers to our research questions. Section \ref{sect:case-study} studies three generated patches in details. Section \ref{sect:discussion} discusses our results and Section \ref{sect:related} presents the related work. Section \ref{sect:conclusion} concludes this paper and proposes future work.

\begin{figure}[!t]
\centering
\includegraphics[width=0.48\textwidth]{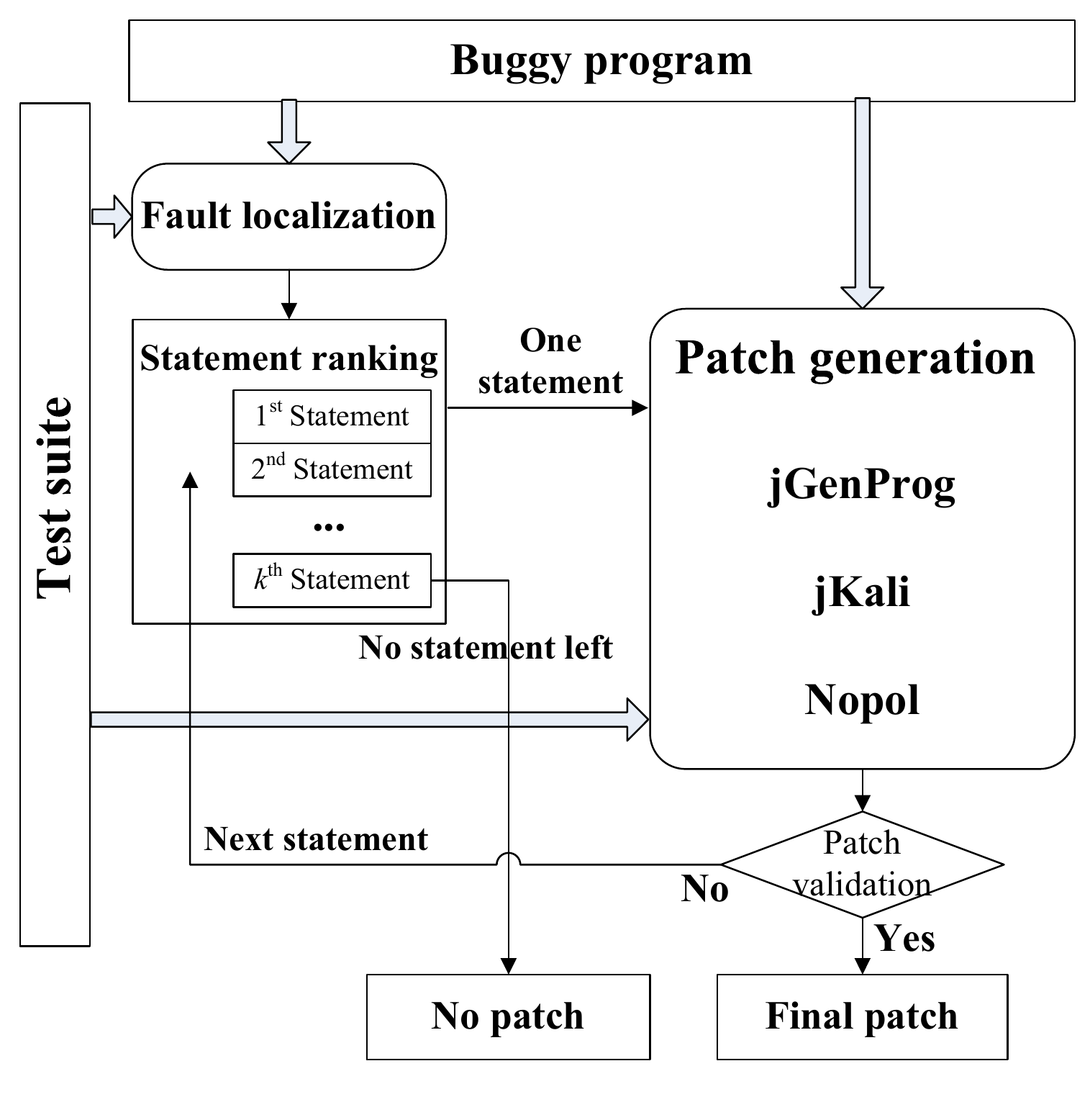}
\caption{Overview of test-suite based repair: it takes a buggy program and its test suite as input, incl. a failing test case; the output is the patch that passes the whole test suite if such patch exists.}
\label{fig:framework}
\end{figure}

\section{Background}
\label{sect:background}

In this paper, we consider one kind of automatic repair called test-suite based repair.
We now give the corresponding background and present the dataset and repair approaches that are used in our experiment.

\subsection{Test-Suite Based Repair}
\label{subsect:automatic-repair}

Test-suite based repair generates a patch according to failing and passing test cases. Different kinds of techniques can be used, such as genetic programming search in GenProg \cite{le2012genprog} and SMT based program synthesis in SemFix \cite{nguyen2013semfix}. Often, before patch generation, a fault localization method is applied to rank the statements according to their suspiciousness. The intuition is that the patch generation technique is more likely to be successful on suspicious statements. 

Fig. \ref{fig:framework} presents a general overview of test-suite based repair approaches. In a repair approach, the input is a buggy program as well as its test suite; the output is a patch that makes the test suite pass, if any. To generate a patch for the buggy program, the executed statements are ranked to identify the most suspicious statements. \textit{Fault localization} is a family of techniques for ranking potential buggy statements \cite{jones2002visualization,abreu2007accuracy,xuan2014test}. Based on the statement ranking, \textit{patch generation} tries to modify a suspicious statement. For instance, GenProg \cite{le2012genprog} adds, removes, and replaces AST nodes. Once a patch is found, the whole test suite is executed to validate the patch; if the patch is not validated by the test suite, the repair approach goes on with next statement and repeats the repair process.

\subsection{Defects4J}
\label{sec:defects4j}

Defects4J by Just et al. \cite{JustJE2014} is a bug database that consists of 357 real-world bugs from five widely-used and large open-source Java projects. Bugs in Defects4J are organized in a unified structure that abstracts over programs, test cases, and patches. 

Defects4J provides research with reproducible software bugs and enables controlled studies in software testing research (e.g., \cite{noor2015test,just2014mutants}). To our knowledge, Defects4J is the largest open database of well-organized real-world Java bugs. In our work, we use four out of five projects, i.e., Commons Lang,\footnote{Apache Commons Lang, \url{http://commons.apache.org/lang}.} JFreeChart,\footnote{JFreeChart, \url{http://jfree.org/jfreechart/}.} Commons Math,\footnote{Apache Commons Math, \url{http://commons.apache.org/math}.} and Joda-Time.\footnote{Joda-Time, \url{http://joda.org/joda-time/}.} We do not use the Closure Compiler project\footnote{Google Closure Compiler, \url{http://code.google.com/closure/compiler/}.} because the test cases in Closure Compiler are organized in a non-conventional way, using scripts rather than standard JUnit test cases. This prevents us from running with our platform and is left for future work.
Table \ref{tab:defects4j} presents the main descriptive statistics of bugs in Defects4J. 

\begin{table}[!t] 
\caption{The Main Descriptive Statistics of Considered Bugs in Defects4J. The Number of Lines of Code and the Number of Test Cases are Extracted from the Most Recent Version of Each Project.}
\label{tab:defects4j}
\centering
\resizebox{0.5\textwidth}{!}{
\setlength\tabcolsep{0.4 ex}
\begin{tabular}{|c|c|c|c|c|}

\hline
Project & \#Bugs & Source KLoC & Test KLoC & \#Test cases \\ \hline\hline
Commons Lang & 65 &  22 & 6 & 2,245 \\ 
JFreeChart & 26 & 96 & 50 & 2,205 \\
Commons Math & 106 & 85 & 19 & 3,602 \\
Joda-Time & 27 & 28 & 53 & 4,130  \\
\hline  \hline
Total & \nbAllBugs & 231 & 128 & 12,182 \\
\hline

\end{tabular}
}
\end{table}

\subsection{Repair Approaches}

\textbf{GenProg} \cite{le2012genprog} repairs programs as follows. It randomly deletes, adds, and replaces abstract syntax tree nodes in the program. The modification point is steered by spectrum based fault localization. Pieces of code that are inserted through addition or replacement always come from the same program, based on the ``redundancy hypothesis'' \cite{DBLP:conf/icse/MartinezWM14}. GenProg is a generic repair approach and does not target any particular fault class.

\textbf{Kali} \cite{qi2015efficient} performs program repair by only removing or skipping code. Even if ``repair'' is achieved in the sense that the patches make the test suite passing, its primary goal is not repair per se. Instead, the goal of Kali is to identify weak test suites and under-specified bugs. 

\textbf{Nopol} \cite{demarco2014automatic,nopoljournal} targets a specific fault class: conditional bugs. It repairs programs by either modifying an existing if-condition or  adding a precondition (aka. a guard) to any statement or block in the code. The modified or inserted condition is synthesized via input-output based code synthesis with SMT \cite{jha2010oracle}.

\section{Experimental Protocol}
\label{sect:protocol}

We present an experimental protocol to assess the effectiveness of different automatic repair approaches on the real-world bugs of Defects4J. The protocol supports the analysis of several dimensions of automatic repair: fixability, patch correctness, under-specified bugs, performance. We first list the Research Questions (RQs) of our work; then we describe the research protocol of our experiment; finally, we present the implementation details. 

\subsection{Research Questions}
\label{subsect:rq}

\subsubsection{\textbf{RQ1}. Fixability} 
Which bugs of Defects4J can be automatically repaired? How many bugs can be repaired by each system?

Fixability is the basic evaluation criterion of automatic repair research. In test-suite based repair, a bug is said to be fixed if the whole test suite passes. To answer this question, we run each repair approach on each buggy program of the dataset under consideration and count the number of bugs, which are patched and can pass the test suite. 

\subsubsection{\textbf{RQ2}. Patch Correctness} 
Which bug fixes are semantically correct (beyond passing the test suite)? 

A patch that passes the whole test suite may not be exactly the same as the patch written by developers. 
It may be syntactically different yet correct.
It may also be incorrect when the test suite is not well-designed and misses important test cases and assertions. 
To answer this question, we manually examine all synthesized patches to identify the correctness as explained in \ref{sec:manual}. 

\subsubsection{\textbf{RQ3}. Under-Specified Bugs} 
Which bugs in Defects4j are not sufficiently specified by the test suite?  

In test-suite based repair, the quality of a synthesized patch is highly dependent on the quality of the test suite. 
In this paper, we define an ``under-specified bug'' as a bug for which the test cases (that specify the expected behavior) have a low coverage and weak assertions.
To find such under-specified bugs, we use two pieces of evidence.
First, we closely look at the results of \jkali.
Since this repair system removes code and skips code execution, if it finds a patch, it hints that a functionality is not specified at all.
Second, for patches found by \jgenprog or Nopol, our manual analysis of the patch may also reveal an under-specification. 

\subsubsection{\textbf{RQ4}. Performance (Execution Time)} 
How long is the execution time of each repair approach? 

It is time-consuming to manually repair a bug. Test-suite based repair automates the process of patch generation. To conduct a quantitative analysis on the performance of automatic repair, we evaluate the execution time of each repair approach.

\subsection{Protocol}

We run three repair systems \jgenprog, \jkali and Nopol on the Defects4J dataset (Section \ref{sec:defects4j}).
Since the experiment requires a large amount of computation, we run it on a grid (Section \ref{sec:grid5000}).
We then manually analyze all the synthesized patches (Section \ref{sec:manual}). 
 
\subsubsection{Repair Systems Under Study}
\label{sec:selected-techniques}

In this experiment, we consider a dataset of bugs in software written in the Java programming language, so we study repair systems that are able to handle this programming language. This is the first selection criterion.
The second one is that it is publicly available.
This left us with Nopol, which comes from our previous work \cite{demarco2014automatic}.
For instance, GenProg \cite{le2012genprog} and Kali \cite{qi2015efficient} only repair C code.
Par \cite{DBLP:conf/icse/KimNSK13} is for Java but not available.

However, we have re-implemented GenProg and Kali for Java software in a repair framework called Astor \cite{astor}.
In the rest of this paper, we will use \jgenprog and \jkali to refer to their re-implementations in Java.
Our motivation of re-implementing GenProg and Kali is the following. 
First, GenProg can be considered as a baseline in the field and is a de-facto point of comparison in the literature. 
Second, Kali is a baseline system to identify under-specified bugs since it consists of only removing and skipping code. For sake of open research and replication, all three systems are made publicly available on Github \cite{githubresults}. 

It can be argued that the results based on a re-implementation do not reflect the actual performance of the original system. 
For instance, a difference in the core algorithms or a bug in the re-implementation may produce invalid empirical results.
In \jgenprog and \jkali, we have carefully followed the description in the corresponding literature. 
We consider that \jkali, our implementation of Kali, is the exact counterpart of the C implementation.
For \jgenprog, we had to make certain decisions on parts due to undefined behaviors or implementation constraints.
In any case, all implementation decisions can be consulted in the source code that is publicly available \cite{githubresults}. 
For future replication, we note that all repair systems and in particular \jgenprog are fully deterministic thanks to the use of a seedable random number generator.

This implementation work represents 9.5K lines of Java code for \jgenprog and \jkali (as measured in Astor), and Nopol has 25K lines of Java code. 

\subsubsection{Large Scale Execution}
\label{sec:grid5000}
We assess three repair approaches on \nbAllBugs bugs. One repair attempt may take hours to be completed. 
Hence, we need a large amount of computation power.
Consequently, we deploy our experiment in Grid'5000, a grid for high performance computing \cite{grid5000}.
In our experiment, we manually set the computing nodes in Grid'5000 to the same hardware architecture. This avoids potential biases of the time cost measurement. 
All the experiments are deployed in the Nancy site of Grid'5000 (located in Nancy, France). 
The cluster management mechanism of Grid'5000 assists our experiments to be reproducible both in fixability and in time cost. 

For each repair approach, we set the timeout to three hours per repair attempt, in order to have a maximum bound on the experiment time (our experiment still takes in total \experimenttimeindays days of computation on Grid'5000).

We stop the execution of a repair attempt after finding the first patch.

\subsubsection{Manual Analysis}
\label{sec:manual}

For correctness assesment, we manually examine the generated patches. 
For each patch, one of the authors (called thereafter an ``analyst'') analyzed the patch correctness, readability, and the difficulty of validating the correctness.

The \textit{correctness} of a patch can be correct, incorrect, or unknown.
The term ``correct'' denotes that a patch is exactly the same or equivalent to the patch that is written by developers.
The equivalence is assessed according to the analyst's understanding of the patch.
Analyzing one patch requires a period between a couple of minutes and several hours of work, depending on the complexity of the synthesized patch. On one hand, a patch that is identical to the one written by developers is obviously true; on the other hand, several patches require a domain expertise that none of the authors has.

The \textit{readability} of the patch can be easy, medium, or hard; and it results from the analyst opinion on the length and complexity of the patch (such as number of variables and method calls used).

The \textit{difficulty} can be easy, medium, hard, or expert. It is related to the effort an analyst carries out for understanding the human patch and the generated patch correctness. For some bugs, it is enough to examine the source code of the patch for determining it correctness, for others the analyst has to debug the buggy and/or the patched application. 
A patch with difficulty ``expert'' means that is impossible for us to validate the correctness due to the required expertise in domain knowledge. 

\section{Empirical Results}
\label{sect:result}

We present and discuss our answers to the research questions that guide this work.
The total execution of the experiment costs \experimenttimeindays days.

\begin{table}[!t]
\caption{Results on the Fixability of \nbAllBugs Bugs in Defects4J with Three Repair Approaches. In Total, the Three Repair Approaches can Repair \nbFixedBugs Bugs (21\%)}.
\label{tab:bug-fix}
\centering
\resizebox{\columnwidth}{.36\textheight}{\setlength\tabcolsep{0.7 ex}
\begin{tabular}{|c|c|c|c|c|}
\hline 
Project & Bug Id & jGenProg & jKali & Nopol   \\
\hline\hline
\multirow{12}{*}{ \rotatebox{90}{JFreeChart}}  
&	C1                	&	 Fixed     	&	 Fixed     	&	 --        	\\
&	C3                	&	 Fixed     	&	 --        	&	 Fixed     	\\
&	C5                	&	 Fixed     	&	 Fixed     	&	 Fixed     	\\
&	C7                	&	 Fixed     	&	 --        	&	 --        	\\
&	C13               	&	 Fixed     	&	 Fixed     	&	 Fixed     	\\
&	C15               	&	 Fixed     	&	 Fixed     	&	 --        	\\
&	C21               	&	 --     	  &	 --     	  &	 Fixed     	\\
&	C25               	&	 Fixed     	&	 Fixed     	&	 Fixed     	\\
&	C26               	&	 --        	&	 Fixed     	&	 Fixed     	\\
\hline
\multirow{7}{*}{ \rotatebox{90}{Commons Lang}} 
&	L39               	&	 --        	&	 --        	&	 Fixed	\\
&	L44               	&	 --        	&	 --        	&	 Fixed     	\\
&	L46               	&	 --        	&	 --        	&	 Fixed	\\
&	L51               	&	 --        	&	 --        	&	 Fixed     	\\
&	L53               	&	 --        	&	 --        	&	 Fixed	\\
&	L55               	&	 --        	&	 --        	&	 Fixed	\\
&	L58               	&	 --        	&	 --        	&	 Fixed     	\\
\hline
\multirow{32}{*}{ \rotatebox{90}{Commons Math}} 
&	M2                	&	 Fixed     	&	 Fixed     	&	 --        	\\
&	M5                	&	 Fixed     	&	 --        	&	 --        	\\
&	M8                	&	 Fixed     	&	 Fixed     	&	 --        	\\
&	M28               	&	 Fixed     	&	 Fixed     	&	 --        	\\
&	M32               	&	 --        	&	 Fixed     	&	 Fixed	\\
&	M33               	&	 --        	&	 --        	&	 Fixed     	\\
&	M40               	&	 Fixed     	&	 Fixed     	&	 Fixed     	\\
&	M42               	&	 --        	&	 --        	&	 Fixed	\\
&	M49               	&	 Fixed     	&	 Fixed     	&	 Fixed     	\\
&	M50               	&	 Fixed     	&	 Fixed     	&	 Fixed     	\\
&	M53               	&	 Fixed     	&	 --        	&	 --        	\\
&	M57               	&	 --        	&	 --        	&	 Fixed     	\\
&	M58               	&	 --        	&	 --        	&	 Fixed     	\\
&	M69               	&	 --        	&	 --        	&	 Fixed	\\
&	M70               	&	 Fixed     	&	 --        	&	 --        	\\
&	M71               	&	 Fixed     	&	 --        	&	 Fixed     	\\
&	M73               	&	 Fixed     	&	 --        	&	 Fixed     	\\
&	M78               	&	 Fixed     	&	 Fixed     	&	 Fixed	\\
&	M80               	&	 Fixed     	&	 Fixed     	&	 Fixed	\\
&	M81               	&	 Fixed     	&	 Fixed     	&	 Fixed     	\\
&	M82               	&	 Fixed     	&	 Fixed     	&	 Fixed	\\
&	M84               	&	 Fixed     	&	 Fixed     	&	 --        	\\
&	M85               	&	 Fixed     	&	 Fixed     	&	 Fixed     	\\
&	M87               	&	 --        	&	 --        	&	 Fixed     	\\
&	M88               	&	 --        	&	 --        	&	 Fixed	\\
&	M95               	&	 Fixed     	&	 Fixed     	&	 --        	\\
&	M97               	&	 --        	&	 --        	&	 Fixed     	\\
&	M104              	&	 --        	&	 --        	&	 Fixed     	\\
&	M105              	&	 --        	&	 --        	&	 Fixed	\\
\hline
\multirow{2}{*}{\rotatebox{90}{\small Time}}
&	T4                	&	 Fixed     	&	 Fixed     	&	 --        	\\
&	T11               	&	 Fixed     	&	 Fixed     	&	 Fixed     	\\
\hline
\hline
Total & 47 (21\%) & 27 (12\%) & 22 (9.8\%) & 35 (15.6\%) \\
\hline 
\end{tabular}
}

\end{table}

\subsection{Fixability} 
\label{subsect:answer-rq1}

\textbf{RQ1}. Which bugs can be automatically repaired? How many bugs can be repaired by each system under study?

The three automatic repair approaches in this experiment are able to together fix \nbFixedBugs bugs of the Defects4J dataset. 
\jgenprog finds a patch for \nbFixedBugsByGenprog bugs; \jkali identifies a patch for \nbFixedBugsByKali bugs; and Nopol synthesizes a condition that makes the test suite passing for \nbFixedBugsByNopol bugs. 
Table \ref{tab:bug-fix} shows the bug identifiers, for which at least one patch is found.
Each line corresponds to one bug in Defects4J and each column denotes the fixability of one repair approach. 
For instance, Bug M2 from Commons Math has been automatically fixed by \jgenprog and \jkali.

As shown in Table \ref{tab:bug-fix}, 
some bugs such as T11 can be fixed by all systems, others by only a single one.
For instance, bug  L39  can only be fixed by Nopol and bug M5 can only be fixed by \jgenprog.
After the controversy about GenProg's effectiveness \cite{qi2015efficient}, it is notable to see that there are bugs for which only \jgenprog works.

Moreover, Table \ref{tab:bug-fix} shows that in project Commons Lang all the bugs are only fixed by Nopol while \jgenprog and \jkali fail to synthesize a single patch. A possible reason is that the program of Commons Lang is more complex than that of Commons Math; both \jgenprog and \jkali cannot handle such a complex search space.

\begin{figure}[!t]
\centering
\includegraphics[width=0.25\textwidth]{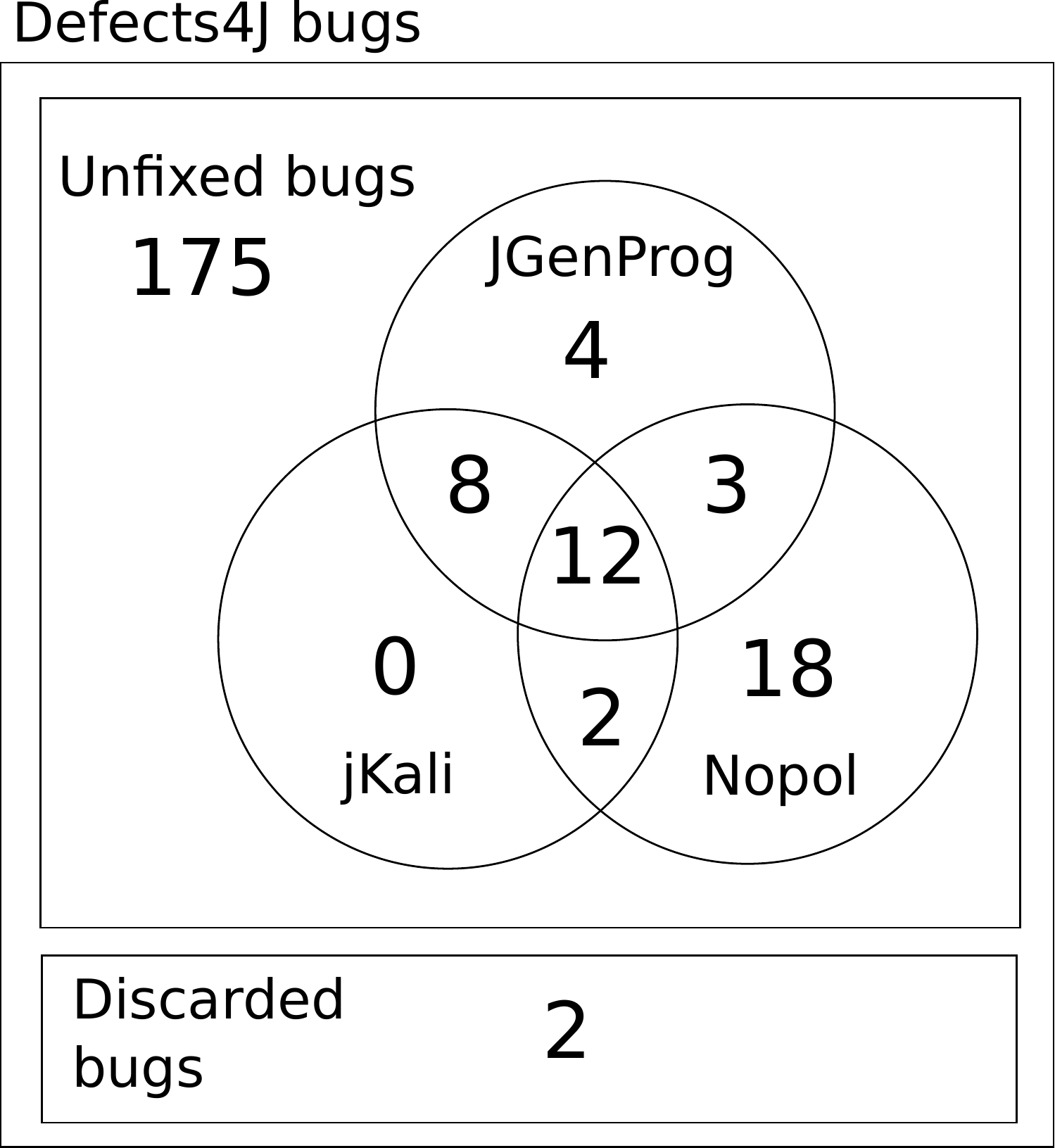}
\caption{Venn diagram that illustrates the bugs commonly fixed by different repair approaches. All fixed bugs by \jkali are also fixed by \jgenprog or Nopol.
}
\label{fig:intersection}
\end{figure}

Fig. \ref{fig:intersection} shows the intersections between the fixed bugs among the three repair approaches as a Venn diagram. Nopol can fix 18 bugs that neither \jgenprog nor \jkali could repair. All the fixed bugs by \jkali can be fixed by \jgenprog or Nopol. For \nbFixedBugsByAll bugs, all three repair systems can generate a patch to pass the test suite.  

To our knowledge, those results are the very first on automatic repair  with the Defects4J benchmark. Recall that they are done with an open-science ethics, all the implementations, experimental code, and results are available on Github \cite{githubresults}. 
Future research in automatic repair may try to fix more bugs than our work. Our experimental framework can be used to facilitate future comparisons by other researchers.   

\noindent\fbox{\parbox{0.47\textwidth}{
\textbf{Answer to RQ1}. In Defects4J, \nbFixedBugs out of \nbAllBugs bugs can be fixed by an automatic repair system. Nopol can fix the largest number of bugs (\nbFixedBugsByNopol bugs). All the fixed bugs by \jkali can be fixed by \jgenprog or Nopol. 
}}

\subsection{Patch Correctness}
\label{subsect:answer-rq2}

\textbf{RQ2}. Which bug fixes are semantically correct (beyond passing the test suite)?  

\begin{table}[!ht]
\caption{The Results of the Manual Assessment of \nbDifferentPatches Patches that are Generated by Three Repair Approaches.}
\label{tab:bug-correct}
\centering
\resizebox{\columnwidth}{.45\textheight}{
\setlength\tabcolsep{0.3ex}
\begin{tabular}{|c|c|c|c|c|c|c|}
\hline 
Project & Bug id & Patch id & Approach & Correctness    & Readability & Difficulty \\
\hline 
\hline 
\multirow{19}{*}{ \rotatebox{90}{JFreeChart}} 
& C1   & 1  & jGenProg & Incorrect         & Easy       & Easy   \\
& C1   & 2  & jKali    & Incorrect         & Easy       & Easy   \\
& C3   & 3  & jGenProg & Unknown           & Medium     & Medium \\
& C3   & 4  & Nopol    & Incorrect         & Easy       & Medium \\
& C5   & 5  & jGenProg & Incorrect         & Easy       & Medium \\
& C5   & 6  & jKali    & Incorrect         & Easy       & Medium \\
& C5   & 7  & Nopol    & \textbf{Correct}  & Easy       & Medium \\
& C7   & 8  & jGenProg & Incorrect         & Easy       & Easy   \\
& C13  & 9  & jGenProg & Incorrect         & Easy       & Easy   \\
& C13  & 10 & jKali    & Incorrect         & Easy       & Easy   \\
& C13  & 11 & Nopol    & Incorrect         & Easy       & Easy   \\
& C15  & 12 & jGenProg & Incorrect         & Easy       & Medium \\
& C15  & 13 & jKali    & Incorrect         & Medium     & Medium \\
& C21  & 14 & Nopol    & Incorrect         & Hard       & Expert \\
& C25  & 15 & jGenProg & Incorrect         & Medium     & Medium \\
& C25  & 16 & jKali    & Incorrect         & Medium     & Medium \\
& C25  & 17 & Nopol    & Incorrect         & Easy       & Easy   \\
& C26  & 18 & jKali    & Incorrect         & Easy       & Medium \\
& C26  & 19 & Nopol    & Incorrect         & Easy       & Medium \\
\hline
\multirow{7}{*}{\rotatebox{90}{Commons Lang}} 
& L39  & 20 & Nopol    & Incorrect         & Easy       & Medium \\
& L44  & 21 & Nopol    & \textbf{Correct}  & Easy       & Medium \\
& L46  & 22 & Nopol    & Incorrect         & Easy       & Medium \\
& L51  & 23 & Nopol    & Incorrect         & Easy       & Easy   \\
& L53  & 24 & Nopol    & Incorrect         & Hard       & Expert \\
& L55  & 25 & Nopol    & \textbf{Correct}  & Easy       & Medium \\
& L58  & 26 & Nopol    & \textbf{Correct}  & Easy       & Medium \\
\hline
\multirow{53}{*}{ \rotatebox{90}{Commons Math}} &
  M2   & 27 & jGenProg & Incorrect         & Easy       & Hard   \\
& M2   & 28 & jKali    & Incorrect         & Easy       & Hard   \\
& M5   & 29 & jGenProg & \textbf{Correct}  & Easy       & Easy   \\
& M8   & 30 & jGenProg & Incorrect         & Easy       & Easy   \\
& M8   & 31 & jKali    & Incorrect         & Easy       & Easy   \\
& M28  & 32 & jGenProg & Incorrect         & Medium     & Hard   \\
& M28  & 33 & jKali    & Incorrect         & Easy       & Hard   \\
& M32  & 34 & jKali    & Incorrect         & Easy       & Easy   \\
& M32  & 35 & Nopol    & Unknown           & Hard       & Expert \\
& M33  & 36 & Nopol    & Incorrect         & Medium     & Medium \\
& M40  & 37 & jGenProg & Incorrect         & Hard       & Hard   \\
& M40  & 38 & jKali    & Incorrect         & Easy       & Medium \\
& M40  & 39 & Nopol    & Unknown           & Hard       & Expert \\
& M42  & 40 & Nopol    & Unknown           & Medium     & Expert \\
& M49  & 41 & jGenProg & Incorrect         & Easy       & Medium \\
& M49  & 42 & jKali    & Incorrect         & Easy       & Medium \\
& M49  & 43 & Nopol    & Incorrect         & Easy       & Medium \\
& M50  & 44 & jGenProg & \textbf{Correct}  & Easy       & Easy   \\
& M50  & 45 & jKali    & \textbf{Correct}  & Easy       & Easy   \\
& M50  & 46 & Nopol    & \textbf{Correct}  & Easy       & Medium \\
& M53  & 47 & jGenProg & \textbf{Correct}  & Easy       & Easy   \\
& M57  & 48 & Nopol    & Incorrect         & Medium     & Medium \\
& M58  & 49 & Nopol    & Incorrect         & Medium     & Hard   \\
& M69  & 50 & Nopol    & Unknown           & Medium     & Expert \\
& M70  & 51 & jGenProg & \textbf{Correct}  & Easy       & Easy   \\
& M71  & 52 & jGenProg & Unknown           & Medium     & Hard   \\
& M71  & 53 & Nopol    & Incorrect         & Medium     & Hard   \\
& M73  & 54 & jGenProg & \textbf{Correct}  & Easy       & Easy   \\
& M73  & 55 & Nopol    & Incorrect         & Easy       & Easy   \\
& M78  & 56 & jGenProg & Unknown           & Easy       & Hard   \\
& M78  & 57 & jKali    & Unknown           & Easy       & Hard   \\
& M78  & 58 & Nopol    & Incorrect         & Medium     & Hard   \\
& M80  & 59 & jGenProg & Incorrect         & Hard       & Medium \\
& M80  & 60 & jKali    & Unknown           & Easy       & Medium \\
& M80  & 61 & Nopol    & Unknown           & Easy       & Medium \\
& M81  & 62 & jGenProg & Incorrect         & Easy       & Medium \\
& M81  & 63 & jKali    & Incorrect         & Easy       & Medium \\
& M81  & 64 & Nopol    & Incorrect         & Easy       & Medium \\
& M82  & 65 & jGenProg & Incorrect         & Easy       & Medium \\
& M82  & 66 & jKali    & Incorrect         & Easy       & Medium \\
& M82  & 67 & Nopol    & Incorrect         & Easy       & Medium \\
& M84  & 68 & jGenProg & Incorrect         & Easy       & Easy   \\
& M84  & 69 & jKali    & Incorrect         & Easy       & Easy   \\     
& M85  & 70 & jGenProg & Unknown           & Easy       & Easy   \\
& M85  & 71 & jKali    & Unknown           & Easy       & Easy   \\
& M85  & 72 & Nopol    & Incorrect         & Easy       & Easy   \\
& M87  & 73 & Nopol    & Incorrect         & Medium     & Expert \\
& M88  & 74 & Nopol    & Incorrect         & Easy       & Medium \\
& M95  & 75 & jGenProg & Incorrect         & Easy       & Hard   \\
& M95  & 76 & jKali    & Incorrect         & Easy       & Hard   \\
& M97  & 77 & Nopol    & Incorrect         & Easy       & Medium \\
& M104 & 78 & Nopol    & Incorrect         & Hard       & Expert \\
& M105 & 79 & Nopol    & Incorrect         & Medium     & Medium \\
\hline
\multirow{5}{*}{\rotatebox{90}{Time}} & 
  T4   & 80 & jGenProg & Incorrect         & Easy       & Medium \\
& T4   & 81 & jKali    & Incorrect         & Easy       & Medium \\
& T11  & 82 & jGenProg & Incorrect         & Easy       & Easy   \\
& T11  & 83 & jKali    & Incorrect         & Easy       & Easy   \\
& T11  & 84 & Nopol    & Incorrect         & Medium     & Medium \\
\hline  
\hline
\multicolumn{4}{|c|}{ \nbDifferentPatches Patches for \nbFixedBugs bugs}        & \nbAnalyzedPatchCorrect Correct        & \nbAnalyzedReaEasy Easy    & \nbAnalyzedDiffHardExpert Hard/Expert  \\
\hline
\multicolumn{7}{|c|}{ 5 patches correct from jGenProg, 1 from jKali and 5 from Nopol}\\
\hline 
\end{tabular}
}
\end{table}

We manually evaluate the correctness of generated patches by the three repair approaches under study as explained in \ref{sec:manual}. In short, a generated patch is considered \textit{correct} if this patch is the same or equivalent as the manually-written patch by developers. A generated patch is \textit{incorrect} if it actually does not completely fix the bug (beyond making the failing test case to pass -- a kind of incomplete bug oracle) or if it breaks an expected behavior (beyond keeping the rest of the test suite passing).

Recall the history of automatic repair research.
It has been hypothesized that a major pitfall of test-suite based repair is that a test suite cannot completely express the program specifications, so it is hazardous to drive the synthesis of a correct patch with a test suite. This comment has been made during conference talks and is common in peer reviews.
Previous works have studied the maintainability of automatic  generated patches \cite{Fry2012} or their aids for debugging task \cite{TaoKKX14}.
However, only recent work by Qi et al. \cite{qi2015efficient} has invested resources to manually analyze the correctness previously-generated patches by test-suite based repair. They found that the vast majority of patches by GenProg in the GenProg benchmark of C bugs are incorrect. 

To answer the question of patch correctness, we have manually analyzed all the patches generated by Nopol, \jgenprog and \jkali in our experiment, \nbDifferentPatches patches in total. 
This represents more than ten full days of work. To our knowledge, only Qi et al. \cite{qi2015efficient} have performed a similar manual assessment of patch synthesized with automatic repair. 
The results of this analysis may be fallible due to the subjective nature of the assessment. For future research, all patches as well as a detailed case study for each of them are made publicly available on Github \cite{githubresults}.  

Table \ref{tab:bug-correct} shows the results of this manual analysis. The ``bug id'' column refers to the Defects4J identifier, while ``Patch id'' is an unique identifier of each patch, for easily identifying the patch on our empirical result page \cite{githubresults}.
The three main columns give the correctness, readability and difficulty as explained in \ref{sec:manual}. 
In total, we have analyzed \nbDifferentPatches patches. Among these patches, \nbFixedBugsByGenprog, \nbFixedBugsByKali, and \nbFixedBugsByNopol patches are synthesized by \jgenprog, \jkali, and Nopol, respectively. 

As shown in Table \ref{tab:bug-correct}, \nbAnalyzedPatchCorrect out of \nbDifferentPatches analyzed patches are correct and \nbAnalyzedIncorrect are incorrect. Meanwhile, for the other \nbAnalyzedUnknown patches, it is not possible to clearly validate the correctness, due to the lack of domain expertise (labeled as \textit{unknown}). Section \ref{sect:case-study} will present three case studies of generated patches via manual analysis.

Among the \nbAnalyzedPatchCorrect correct patches, \jgenprog, \jkali, and Nopol contribute to 5, 1, and 5 patches, respectively. All the correct patches by \jgenprog and \jkali come from Commons Math; 3 correct patches by Nopol come from Commons Lang, one comes from JFreeChart and the other from Commons Math. 

For the incorrect patches, the main reasons are as follows.
First, all three approaches are able to remove some code (pure removal for \jkali, replacement for \jgenprog, precondition addition for Nopol). The corresponding patches simply exploit some under-specification and remove the faulty but otherwise not used behavior. This goes along the line of Qi et al.'s results \cite{qi2015efficient}. 
When the expected behavior seems to be well-specified (according to our understanding of the domain), the incorrect patches tend to overfit to the test data. For instance, if a failing test case handles a $2 \times 2$ matrix, 
the patch may use such test data to incorrectly force the patch to be suitable only for matrices of size of $2 \times 2$. This overfitting characteristic has recently been studied by Smith and colleagues \cite{Smith15fse}.

Among \nbDifferentPatches analyzed patches, \nbAnalyzedReaEasy patches are identified as easy to read and understand.
For the difficulty of patch validation, \nbAnalyzedDiffHardExpert patches are labeled as hard or expert. This result shows that it is hard and time consuming to conduct the validation of patches. 

Overall, our experimental results confirm the conclusion of Qi et al. \cite{qi2015efficient} about incorrect patches due to under-specification: most patches found by test-suite based repair are incorrect. This confirmation is two-sided. 
First, both results by Qi et al. and by us have the same conclusion, but come from different bug benchmarks. 
Second, the finding holds for different systems: while  Qi et al.'s results were made on GenProg, the same finding holds for Nopol. 

This leads to two directions for future work. 
First, test case generation and test suite amplification may be able to reduce the risk that the synthesized patches overfit the test case data. Second, we imagine that different repair algorithms may be more or less subject to overfitting.

\noindent\fbox{\parbox{0.47\textwidth}{
\textbf{Answer to RQ2}. Based on manual examination of patch correctness, we find out that only \nbAnalyzedPatchCorrect out of \nbDifferentPatches generated patches are semantically correct. The repair systems under study tend to suffer from  weak test suite. There exists large room for improving the effectiveness of test-suite based repair. }}

\subsection{Under-specified bugs}
\label{subsect:answer-rq3}
\textbf{RQ3}. Which bugs in Defects4j are not sufficiently specified by the test suite?

As shown in Section \ref{subsect:answer-rq1}, the repair system \jkali can generate a patch for \nbFixedBugsByKali bugs. Among these generated patches, from our manual evaluation, we find out that 18 patches are incorrect (other 3 patches are unknown). In each of those generated patches by \jkali, one statement is removed or skipped to eliminate the failing program behavior, instead of making it correct. 
This kind of patches shows that the corresponding test suite is too weak with respect to the buggy functionality. 
The assertions that specify the expected behavior of the removed statement and the surrounding code are inexistent or too weak. 

One exception among \nbFixedBugsByKali patches by \jkali is the patch of Bug M50. As shown in Section  \ref{subsect:answer-rq2}, the patch of Bug M50 is correct. That is, the statement removal is the correct patch. 
Another special case is Bug C5 which is patched by \jkali (incorrect) and by Nopol (correct). The latter approach produces a patch similar to that one done by the developer. 
A patch (written by developer or automatically generated) that fixes an under-specifier bug could introduce new bugs (studied previously by Gu et al. \cite{Zhongxian2010buggyfix}) or it could not be completely correct due to a weak test suite used as bug oracle \cite{qi2015efficient}.
Table \ref{tab:bug-remove} summarizes this finding and list the under-specified bugs.

This result is important for future research on automatic repair with Defects4J.
First, any repair system that claims to correctly fix one of those bugs should be validated with a detailed manual analysis of patch correctness, to check whether the patch is not a variation on the trivial removal solution.
Second, those bugs can be considered as the most challenging ones of Defects4J. To fix them, a repair system must somehow reason on the expected functionality below what is encoded in the test suite.
This is what was actually been done by the human developer. 
A repair system that is able to produce a correct patch for those bugs would be a great advance for the field.

\noindent\fbox{\parbox{0.47\textwidth}{
\textbf{Answer to RQ3}. There are under-specified bugs in the Defects4J dataset. For them, the test suite does not accurately specify the expected behavior and can be trivially repaired by removing code.
To us, they are the most challenging bugs: to automatically repair them, one needs to reason on the expected functionality below what is encoded in the test suite, to take into account a source of information other than the test suite execution.}}

\begin{table}[!t] 
\caption{The Most Challenging Bugs of Defects4J Because of Under-specification.}
\label{tab:bug-remove}
\centering
\begin{tabular}{|c|c|}

\hline
Project      & Bug ID    \\ \hline\hline
Commons Math & M2, M8,M28,M32,M40, M49, M78, \\ 
	    	     & M80, M81, M82,M84, M85,M95 \\  \hline
JFreeChart & C1,C5, C13, C15, C25,C26\\\hline
Time &T4,T11\\\hline
\end{tabular}
\end{table}

\subsection{Performance}
\label{subsect:answer-rq4}
\textbf{RQ4}. How long is the execution time for each repair approach on one bug? 

For real applicability in industry, automatic repair approaches must execute fast enough. 
By ``fast enough'', we mean an acceptable time period, which depends on the usage scenario of automatic repair and on the hardware. For instance, if automatic repair is meant to be done in the IDE, repair time should last at most some minutes on a standard desktop machine. 
On the contrary, if automatic repair is meant to be done  on a continuous integration server, it is acceptable to last hours on a more powerful server hardware configuration. 

The experiments in this paper are run on a grid where most of nodes have comparable characteristics. 
Typically, we use machines with Intel Xeon X3440 Quad-core processor and 15GB RAM.
Table \ref{tab:bug-time} shows the time cost of patch generation in hours for bugs without timeout. As shown in Table \ref{tab:bug-time}, the median time for one bug by \jgenprog is around one hour.
The fastest repair attempt yields a patch in 31 seconds (for Nopol). 
The median time to synthesize a patch is 6.7 minutes. 
This means that the execution time of automatic repair approaches is comparable to the time of manual repair by developers. It may be even faster, but we don't know the actual repair time by real developers for the bug of the dataset.

When a repair exists, it is found within minutes. This means that most of the time of the \experimenttimeindays days of computation for the experiment is spent on unfixed bugs, which reach the timeout.
For \jgenprog, it is always the case, because the search space is extremely large.
For \jkali, we often completely explore the search space, and we only reach the timeout in 20 cases.
For Nopol, the timeout is reached in 26 cases, either due to the search space of covered statements or the SMT synthesis that becomes slow in certain cases.
One question is whether a larger timeout would improve the effectiveness. 
According to this experiment, the answer is no. The repairability is quite binary: either a patch is found fast, or the patch cannot be found at all. This preliminary observation calls for future research.

\begin{table}[!t] 
\caption{Time Cost of Patch Generation}
\label{tab:bug-time}
\centering
\resizebox{0.40\textwidth}{!}{
\begin{tabular}{|c|c|c|c|}

\hline
Time cost& \jgenprog  & \jkali       & Nopol        \\\hline\hline
Min    & 40 sec       & 36 sec       & 31sec        \\
Median & 1h 01m       & 18m 45sec    & 22m 30sec      \\
Max    & 1h 16m       & 1h 27m       & 1h 54m       \\\hline
Average& 55m 50sec    & 23m 33sec    & 30m 53sec    \\
Total  & 8 days 12h   & 3 days 6h    & 7 days 3h    \\\hline
\end{tabular}
}
\end{table}

\noindent\fbox{\parbox{0.47\textwidth}{
\textbf{Answer to RQ4}. 
For real bugs on large Java projects, the average repair time of the three systems under study is resp. 23, 30 and 55 minutes.
This means that the performance time of the considered repair systems is within reach of practical applicability.}}

\subsection{Other Findings in Defects4J}
\label{subsect:other-findings}

Our manual analysis of results enables us to uncover two problems in Defects4J.
First, we found that bug \#8 from project JFreeChart (C8) is flaky, which depends on the machine configuration.
Second, bug \#99 from Commons Math (M99) is identical to bug M97. Both issues were reported to the authors of Defects4J and will be solved in future releases of Defects4J.

\section{Case Studies}
\label{sect:case-study}

In this section, we present three case studies of generated patches by \jgenprog, \jkali, and Nopol, respectively. These case studies are pieces of evidence that:
\begin{itemize}
  \item Automatic repair is able to find correct patches (Sections \ref{subsect:case-m70} and \ref{subsect:case-l55}), but also fails with incorrect patches (Section \ref{subsect:case-m8}). 
  \item It is possible to automatically generate the same patch as the manual patch written by the developer (Section \ref{subsect:case-m70}). 
  \item To pass the whole test suite, an automatic repair approach may generate useless patches (Section \ref{subsect:case-m8}). 
\end{itemize}

\subsection{Case Study of M70, Bug that is Only Fixed by \jgenprog}
\label{subsect:case-m70}

In this section, we study Bug M70, which is fixed by \jgenprog, but cannot be fixed by \jkali and Nopol.  

Bug M70 in Commons Math is about univariate real function analysis. 
Fig. \ref{fig:case-m70} presents the buggy method of Bug M70. This buggy method contains only one statement, a method call to an overloaded method. In order to perform the correct calculation, the call has to be done to a with an additional parameter \mycode{UnivariateRealFunction f} (at Line 1) to the method call. Both the manually-written patch and the patch by \jgenprog add the parameter \mycode{f} to the method call (at Line 5). 
This patch generated by \jgenprog is considered correct since the it is the same as that by developers. 

\begin{figure}[!t]
\centering
\noindent\begin{minipage}{0.4\textwidth}
\begin{lstlisting}[numbers=left]
double solve(UnivariateRealFunction f, 
      double min, double max, double initial)
    throws MaxIterationsExceededException, 
      FunctionEvaluationException {
// FIX: return solve(f, min, max);
  return solve(min, max); 
}
\end{lstlisting}

\end{minipage}
\caption{Code snippet of Bug M70. The manually-written patch and the patch by \jgenprog are the same, which is shown in the \mycode{FIX} comment at Line 5, which adds a parameter to the method call.} 
\label{fig:case-m70}
\end{figure}

To fix Bug M70, \jgenprog generates a patch by replacing the method call by another one, which is picked elsewhere in the same class. This bug cannot be fixed by either \jkali or Nopol. \jkali removes and skips statements; Nopol only handles bugs that are related to \ourif conditions. Indeed, the fact that certain bugs are only fixed by one tool confirms that the fault classes addressed by each approach are not identical. 
\textbf{To sum up, Bug M7 shows that the GenProg algorithm, as implemented in \jgenprog, is capable of uniquely repairing real Java bugs (only GenProg succeeds).}

\subsection{Case Study of M8, Bug that is Incorrectly Fixed by \jkali and \jgenprog}
\label{subsect:case-m8} 

In this section, we present a case study of Bug M8, which is fixed by \jkali as well as \jgenprog, but fails to be fixed by Nopol.  

\begin{figure}[!t]
\centering
\noindent\begin{minipage}{0.4\textwidth}
\begin{lstlisting}[numbers=left]
T[] sample(int sampleSize)  {
  if (sampleSize <= 0) {
    throw new NotStrictlyPositiveException([...]);
  }
// MANUAL FIX: 
// Object[] out = new Object[sampleSize];
  T[] out = (T[]) Array.newInstance(
    singletons.get(0).getClass(), sampleSize);
  for (int i = 0; i < sampleSize; i++) {
// FIX: removing the following line
    out[i] = sample();  
  }
  return out;
}
\end{lstlisting}

\end{minipage}
\caption{Code snippet of Bug M8. The manually-written patch is shown in the \mycode{MANUAL FIX} comment at Lines 5 and 6 (changing a variable type). The patch by \jkali in the \mycode{FIX} comment removes the loop body at Line 11. }
\label{fig:case-m8}
\end{figure}

Bug M8\footnote{Bug ID in the bug tracking system of Commons Math is Math-942, \url{http://issues.apache.org/jira/browse/MATH-942}.} in Commons Math, is about the failure to create an array of a random sample from a discrete distribution.
Listing \ref{fig:case-m8} shows an excerpt of the buggy code and the corresponding manual and synthesized fixes (from \mycode{class DiscreteDistribution<T>}).
The method \mycode{sample} receives the expected number \mycode{sampleSize} of random values and returns an array of the type \mycode{T[]}. 

The bug is due to an exception thrown at line 11 during the assignment to \mycode{out[i]}.
The method \mycode{Array.newInstance(class, int)} requires a class of a data type as the first parameter. The bug occurs when 
a) the first parameter is of type \mycode{T1}, which is a sub-class of \mycode{T}  
and 
b) one of the samples is an object which is of type \mycode{T2}, which is a sub-class of \mycode{T}, but not of type \mycode{T1}. 
Due to the incompatibility of types T1 and T2, an \mycode{ArrayStoreException} is thrown when this object is assigned to the array.  

In the manual patch, the developers change the array type in its declaration (from \mycode{T[]} to \mycode{Object[]}) and the way the array is instantiated.
The patch generated by \jkali simply removes the statement, which assigns \mycode{sample()} to the array.
As consequence, method \mycode{sample} never throws an exception but returns an empty array (only containing null values).
This patch passes the failing test case and the full test suite as well.
The reason of this is that the test case has only one assertion: it asserts that the array size is equal to 1. There is no assertion on the content of the returned array.
However, despite passing the test suite, the patch is clearly incorrect. 
This is an example of a bug that is not well specified by the test suite. 
For this bug, \jgenprog can also generate a patch by replacing the assignment by a side-effect free statement, which is semantically equivalent to removing the code. 
\textbf{To sum up, Bug M8 is an archetypal example of  under-specified bugs as detected by the \jkali system.}

\subsection{Case Study of L55, Bug that is Fixed by Nopol, Equivalent to the Manual Patch}
\label{subsect:case-l55} 

In this section, we present a case study of Bug L55, which is only fixed by Nopol, but cannot be fixed by \jgenprog or \jkali.
Recall that Nopol \cite{demarco2014automatic} focuses on condition-related bugs. 

\begin{figure}[!t]
\centering
\noindent\begin{minipage}{0.4\textwidth}
\begin{lstlisting}[numbers=left]
void stop() {
  if (this.runningState != STATE_RUNNING 
      && this.runningState != STATE_SUSPENDED) {
    throw new IllegalStateException(...);
  }
// MANUAL FIX: 
// if (this.runningState == STATE_RUNNING)
// NOPOL FIX: 
// if (stopTime < StopWatch.STATE_RUNNING)
  stopTime = System.currentTimeMillis();
  this.runningState = STATE_STOPPED;
}
\end{lstlisting}

\end{minipage}
\caption{Code snippet of Bug L55. The manually-written patch is shown in the \mycode{MANUAL FIX} comment at Lines 6 and 7 while the patch by Nopol is shown in the \mycode{NOPOL FIX} at Lines 8 and 9. The patch by Nopol is equivalent to the manually-written patch by developers.}
\label{fig:case-l55}
\end{figure}

Bug L55 in Commons Lang relates a utility class for timing. The bug appears when the user stops a suspended timer: the stop time saved by the suspend action is overwritten by the stop action. Fig. \ref{fig:case-l55} presents the buggy method of Bug L55. In order to solve this problem, the assignment at Line 10 has to be done only if the timer state is running.

As shown in Fig. \ref{fig:case-l55}, the manually-written patch by the developer adds a precondition before the assignment at Line 10 and it checks that the current timer state is running (at Line 7). 
The patch by Nopol is different from the manually-written one. The Nopol patch compares the stop time variable to a integer constant (at Line 9), which is pre-defined in the program class and equals to $1$. In fact, when the timer is running, the stop time variable is equals to $-1$; when it is suspended, the stop time variable contains the stop time in millisecond. Consequently, both preconditions by developers and by Nopol are equivalent and correct. Despite being equivalent, the manual patch remains more understandable.
This bug is neither fixed by \jgenprog nor \jkali. To our knowledge, Nopol is the only approach that contains a strategy of adding preconditions to original statements, which does not exist in \jgenprog or \jkali. 
\textbf{To sum up, Bug L55 shows an example of a repaired bug, 1) that is in a hard-to-repair project (only Nopol succeeds) and 2) whose specification by the test suite is good enough to drive the synthesis of a correct patch.}

\subsection{Summary}. In this section, we have presented detailed case studies of three patches that are automatically generated  for three real-world bugs of Defects4J. Our case studies show that automatic repair approaches are able to fix real bugs. However, different factors, in particular the weakness of some test cases, yield clearly incorrect patches.

\section{Discussion}
\label{sect:discussion} 

\subsection{Threats to Validity}

\textbf{Implementations of GenProg and Kali}. In \jgenprog and \jkali, we have re-implemented the GenProg and Kali algorithms in Java, according to the related papers. Although we have tried our best to understand and implement these two approaches, there still exists a threat that our implementations do not exactly produce the same results as the original systems would. Since GenProg and Kali are not written for Java, the re-implementation was the only way to conduct a comparison. To find re-implementation issues, our systems are publicly available on Github.

\textbf{Bias of assessing the correctness, readability, and difficulty}. In our work, each patch in Table \ref{tab:bug-correct} is validated by an analyst, which is one of the authors. An analyst manually identifies the correctness of a patch and labels the related readability and difficulty. However, it may happen that the judgment by analysts is incorrect. In our experiment, since manual analysis is very tedious, we did not cross analysis (more than one analyst per patch). However, we share our results online on the experiment Github repository to let readers have a preliminary idea of the difficulty of our analysis work and the correctness of generated patches (see Section \ref{subsect:answer-rq1}). For assessing the equivalence, one solution would be to use automatic technique, as done in mutation testing. However, whether the current equivalence detection techniques scale on large real Java code is an open question.

\textbf{Random nature of \jgenprog}. \jgenprog, as the original GenProg implementation, has a random component. Statements and mutations are randomly chosen based on rules during the search. Consequently, it may happen that a different run of \jgenprog with the same timeout would repair more bugs. Due to the ultra large computation time of the experiment, it was impossible for us to run \jgenprog enough times to assess this. We leave this to future work.

\textbf{Presence of multiple patches}. 
In this experiment, we stop the execution of a repair system once the first patch is found. This is the patch that is manually analyzed. 
However, as also experienced by others, there are often several if not dozens of different patches per bug. 
It might happen that a correct patch lies somewhere in this set of generated patches. We did not manually analyze all generated patches because it would require months of manual work. This finding shows that there is a need for research on approaches that order the generated patches so as to reveal the most likely to be correct.

\subsection{Impact of Flaky Tests on Repair}
Our experiment uncovered one flaky test in Defects4J (C8).
We realized that flaky tests have a direct impact on automatic repair.
If the failing test case is flaky, the repair system might conclude that a correct patch has been found while it is actually correct.
If one of the passing test cases is flaky, the repair system might conclude that a patch has introduced a regression while it is not the case, this results in an underestimation of the effectiveness of the repair technique.

\subsection{Reflections on GenProg}

The largest evaluations of GenProg are by Le~Goues et al. \cite{DBLP:conf/icse/GouesDFW12} and Qi et al. \cite{qi2015efficient}.
The former reports that 55/105 bugs are fixed (under the definition that the patch passes the test suite), while the latter argued that only 2/105 bugs are correctly fixed (under the definition that the patch passes the test suite and that the patch is correct and acceptable from the viewpoint of the developer). The difference is due to an experimental error and the presence of under-specified bugs.

In this paper, we find that our re-implementation of GenProg, \jgenprog correctly fixes \nbAnalyzedGenprogCorrect /\nbAllBugs (2.2\%) bugs. In addition, it uniquely fixes 4 bugs, such as M70 discussed in Section \ref{subsect:case-m70}.

We interpret those results as follows.
First, having correctly and uniquely fixed bugs indicates that the core intuition of GenProg is valid, and that GenProg can be a component of an integrated repair tool that would mix different repair techniques.
Second, the difference in repair rate is probably due to the inclusion criteria of both benchmarks (GenProg and Defects4J). To our opinion, none of them reflect the actual distribution of all bug kinds and difficulty in nature.
Also, one factor may be the programming language under repair. One could hypothesize that GenProg is better suited to fix procedural code (as C is), whereas the complexity of OO code written in Java does not lie in the control flow of the methods, but in the class design and interactions. New experiments have to be designed to validate or invalidate this hypothesis.

\section{Related Work}
\label{sect:related}

\subsection{Real-World Datasets of Bugs}
\label{subsect:related-dataset}

The academic community has set up real-world bug data to evaluate their software testing methods and to analyze their performance in practice. For instance, Do et al. \cite{do2005supporting} propose a controlled experimentation platform for testing techniques. Their dataset is included in SIR database, which provides a widely-used testbed in debugging and test suite optimization.

Dallmeier et al. \cite{dallmeier2007extraction} propose iBugs, a benchmark for bug localization obtained by extracting historical bug data.
BugBench by Lu et al. \cite{lu2005bugbench} and BegBunch by Cifuentes et al. \cite{cifuentes2009begbunch} are two benchmarks that have been built to evaluate bug detection tools. 
The PROMISE repository \cite{promise2015} is a collection of datasets in various fields of software engineering. Le~Goues et al. \cite{LeGoues15tse} have designed a benchmark of C bugs which is an extension of the GenProg benchmark.

In this experience report, we employ Defects4J by Just et al. \cite{JustJE2014} to evaluate software repair. This database includes well-organized programs, bugs, and their test suites. The bug data in Defects4J has been extracted from the recent history of five widely-used Java projects. 
To us, Defects4J is the best dataset of real world Java bugs, both in terms of size and quality.
To our knowledge, our experiment is the first that evaluates automatic repair techniques via Defects4J.

\subsection{Test-Suite Based Repair Approaches}
\label{subsect:related-repair}

The idea of applying evolutionary optimization to repair derives from Arcuri \& Yao \cite{DBLP:conf/cec/ArcuriY08}. Their work applies co-evolutionary computation to automatically generate bug fixes. GenProg by Le Goues et al. \cite{le2012genprog} applies genetic programming to the AST of a buggy program and generates patches by adding, deleting, or replacing AST nodes. PAR by Kim et al. \cite{DBLP:conf/icse/KimNSK13} leverages patch patterns learned from human-written patches to find readable patches. RSRepair by Qi et al. \cite{qi2014strength} uses random search instead of genetic programming. This work shows that random search is more efficient in finding patches than genetic programming. Their follow-up work \cite{DBLP:conf/icsm/QiML13} uses test case prioritization  to reduce the cost of patch generation.

Debroy \& Wong \cite{DBLP:conf/icst/DebroyW10} propose a mutation-based repair method inspired from mutation testing. This work combines fault localization with program mutation to exhaustively explore a space of possible patches. Kali by Qi et al. \cite{qi2015efficient} has recently been proposed to examine the fixability power of simple actions, such as statement removal. 

SemFix by Nguyen et al. \cite{nguyen2013semfix} is a notable constraint based repair approach. This approach provides patches for assignments and conditions by combining symbolic execution and code synthesis. Nopol by DeMarco et al. \cite{demarco2014automatic} is also a constraint based method, which focuses on fixing bugs in \ourif conditions and missing preconditions. DirectFix by Mechtaev et al.  \cite{mechtaev2015directfix} achieves the simplicity of patch generation with a Maximum Satisfiability (MaxSAT) solver to find the most concise patches. Relifix \cite{relifix} focuses on regression bugs. SPR \cite{Long15} defines a set of staged repair operators so as to early discard  many candidate repairs that cannot pass the supplied test suite and eventually to exhaustively explore a small and valuable search space.

Besides test-suite based repair, other repair setups have been proposed.  Yu et al. \cite{DBLP:journals/tse/0001FNWMZ14}  
proposed a contract based method for automatic repair. Other related repair methods include atomicity-violation fixing (e.g. \cite{DBLP:conf/pldi/JinSZLL11}), runtime error repair (e.g. \cite{DBLP:conf/pldi/LongSR14}), and domain-specific repair (e.g. \cite{DBLP:conf/icse/SamirniSAMTH12,gopinath2014data}).

\subsection{Empirical Investigation of Automatic Repair}
\label{subsect:related-analysis}

Beyond proposing new repair techniques, there is a thread of research on empirically investigating the foundations, impact and applicability of automatic repair. 

On the goodness of synthesized patches, 
 Fry et al. \cite{DBLP:conf/issta/FryLW12} conducted a study of machine-generated patches based on 150 participants and 32 real-world defects. Their work shows that machine-generated patches are slightly less maintainable than human-written ones. Tao et al. \cite{TaoKKX14} performed a similar study to study whether machine-generated patches assist human debugging. Monperrus \cite{monperrus2014critical} discussed in depth the acceptability criteria of synthesized patches. 

Martinez \& Monperrus \cite{DBLP:journals/ese/MartinezM15} studied thousands of commits to mine repair models from manually-written patches. They later investigated \cite{DBLP:conf/icse/MartinezWM14} the redundancy assumption in automatic repair (whether you can fix bugs by rearranging existing code). 
Zhong \& Su \cite{zhong2015an} conducted a case study on over 9,000 real-world patches and found two important facts for automatic repair: for instance, their analysis outlines that some bugs are repaired with changing the configuration files.

\section{Conclusion}
\label{sect:conclusion} 

We have presented an experience report of a large scale evaluation of automatic repair approaches. Our experiment was conducted with three automatic repair systems on \nbAllBugs bugs in the Defects4J dataset. We find out that the systems under consideration can synthesize a patch for \nbFixedBugs out of \nbAllBugs. 
\emph{Since the dataset only contains real bugs from large-scale Java software, this is a piece of evidence about the applicability of automatic repair in practice.}

Our findings indicate that there is a need for better repair algorithms, for instance, better code synthesis with multiple method calls or more complex patches applied at multiple locations. This may ``unlock'' other bugs from Defects4J.
Also, we suggest that the presence of multiple patches indicate a need for research on approaches that order the generated patches so as to reveal the most likely to be correct. Preliminary work is being done on this topic \cite{prophet}.

To our opinion, there is also a need for research on test suites. 
For instance, any approach that automatically enriches test suites with new tests \cite{Shamshiri2015} or stronger assertions would have a direct impact on repair, by preventing the synthesis of incorrect patches. 

The three repair systems and all experimental results are publicly available at \\ \url{http://github.com/Spirals-Team/defects4j-repair/}.

\newpage
\bibliographystyle{abbrv}
\balance
\bibliography{IEEEabrv,references}

\begin{thebibliography}{10}

\bibitem{promise2015}
Promise repository.
\newblock \url{http://openscience.us/repo/}.
\newblock Accessed: 2015-05-01.

\bibitem{githubresults}
The github repository of the experimental data.
\newblock \url{https://github.com/Spirals-Team/defects4j-repair}, 2015.

\bibitem{abreu2007accuracy}
R.~Abreu, P.~Zoeteweij, and A.~J. Van~Gemund.
\newblock On the accuracy of spectrum-based fault localization.
\newblock In {\em Testing: Academic and Industrial Conference Practice and
  Research Techniques-MUTATION, 2007. TAICPART-MUTATION 2007}, pages 89--98.
  IEEE, 2007.

\bibitem{DBLP:conf/cec/ArcuriY08}
A.~Arcuri and X.~Yao.
\newblock A novel co-evolutionary approach to automatic software bug fixing.
\newblock In {\em Proceedings of the {IEEE} Congress on Evolutionary
  Computation}, pages 162--168, 2008.

\bibitem{grid5000}
R.~Bolze, F.~Cappello, E.~Caron, M.~Dayd{\'e}, F.~Desprez, E.~Jeannot,
  Y.~J{\'e}gou, S.~Lanteri, J.~Leduc, N.~Melab, et~al.
\newblock Grid'5000: a large scale and highly reconfigurable experimental grid
  testbed.
\newblock volume~20, pages 481--494. SAGE Publications, 2006.

\bibitem{cifuentes2009begbunch}
C.~Cifuentes, C.~Hoermann, N.~Keynes, L.~Li, S.~Long, E.~Mealy, M.~Mounteney,
  and B.~Scholz.
\newblock Begbunch: Benchmarking for c bug detection tools.
\newblock In {\em Proceedings of ISSTA}, pages 16--20, New York, USA, 2009.
  ACM.

\bibitem{dallmeier2007extraction}
V.~Dallmeier and T.~Zimmermann.
\newblock {Extraction of Bug Localization Benchmarks from History}.
\newblock In {\em Proceedings of the Twenty-second IEEE/ACM International
  Conference on Automated Software Engineering}, pages 433--436, 2007.

\bibitem{DBLP:conf/icst/DebroyW10}
V.~Debroy and W.~E. Wong.
\newblock Using mutation to automatically suggest fixes for faulty programs.
\newblock In {\em Third International Conference on Software Testing,
  Verification and Validation, {ICST} 2010, Paris, France, April 7-9, 2010},
  pages 65--74, 2010.

\bibitem{demarco2014automatic}
F.~Demarco, J.~Xuan, D.~L. Berre, and M.~Monperrus.
\newblock Automatic repair of buggy if conditions and missing preconditions
  with smt.
\newblock In {\em Proceedings of the 6th International Workshop on Constraints
  in Software Testing, Verification, and Analysis}, pages 30--39. ACM, 2014.

\bibitem{do2005supporting}
H.~Do, S.~Elbaum, and G.~Rothermel.
\newblock Supporting controlled experimentation with testing techniques: An
  infrastructure and its potential impact.
\newblock {\em Empirical Software Engineering}, 10(4):405--435, 2005.

\bibitem{prophet}
FanLong and M.~Rinard.
\newblock {Prophet: Automatic Patch Generation via Learning from Successful
  Human Patches}.
\newblock Technical report, {MIT}, 2015.

\bibitem{Fry2012}
Z.~P. Fry, B.~Landau, and W.~Weimer.
\newblock A human study of patch maintainability.
\newblock In {\em Proceedings of the 2012 International Symposium on Software
  Testing and Analysis}, ISSTA 2012, pages 177--187, New York, NY, USA, 2012.
  ACM.

\bibitem{DBLP:conf/issta/FryLW12}
Z.~P. Fry, B.~Landau, and W.~Weimer.
\newblock A human study of patch maintainability.
\newblock In {\em Proceedings of the International Symposium on Software
  Testing and Analysis}, pages 177--187, 2012.

\bibitem{gopinath2014data}
D.~Gopinath, S.~Khurshid, D.~Saha, and S.~Chandra.
\newblock Data-guided repair of selection statements.
\newblock In {\em Proceedings of the 36th International Conference on Software
  Engineering}, pages 243--253. ACM, 2014.

\bibitem{Zhongxian2010buggyfix}
Z.~Gu, E.~Barr, D.~Hamilton, and Z.~Su.
\newblock Has the bug really been fixed?
\newblock In {\em Software Engineering, 2010 ACM/IEEE 32nd International
  Conference on}, volume~1, pages 55--64, May 2010.

\bibitem{jha2010oracle}
S.~Jha, S.~Gulwani, S.~A. Seshia, and A.~Tiwari.
\newblock Oracle-guided component-based program synthesis.
\newblock In {\em Proceedings of the International Conference on Software
  Engineering}, volume~1, pages 215--224. IEEE, 2010.

\bibitem{DBLP:conf/pldi/JinSZLL11}
G.~Jin, L.~Song, W.~Zhang, S.~Lu, and B.~Liblit.
\newblock Automated atomicity-violation fixing.
\newblock In {\em Proceedings of the 32nd {ACM} {SIGPLAN} Conference on
  Programming Language Design and Implementation}, pages 389--400, 2011.

\bibitem{jones2002visualization}
J.~A. Jones, M.~J. Harrold, and J.~Stasko.
\newblock Visualization of test information to assist fault localization.
\newblock In {\em Proceedings of the 24th international conference on Software
  engineering}, pages 467--477. ACM, 2002.

\bibitem{JustJE2014}
R.~Just, D.~Jalali, and M.~D. Ernst.
\newblock {Defects4J}: A database of existing faults to enable controlled
  testing studies for {J}ava programs.
\newblock In {\em Proceedings of the International Symposium on Software
  Testing and Analysis (ISSTA)}, pages 437--440, July~23--25 2014.

\bibitem{just2014mutants}
R.~Just, D.~Jalali, L.~Inozemtseva, M.~D. Ernst, R.~Holmes, and G.~Fraser.
\newblock Are mutants a valid substitute for real faults in software testing.
\newblock In {\em 22nd International Symposium on the Foundations of Software
  Engineering (FSE 2014)}, 2014.

\bibitem{DBLP:conf/icse/KimNSK13}
D.~Kim, J.~Nam, J.~Song, and S.~Kim.
\newblock Automatic patch generation learned from human-written patches.
\newblock In {\em Proceedings of the 35th International Conference on Software
  Engineering}, pages 802--811, 2013.

\bibitem{le2012systematic}
C.~Le~Goues, M.~Dewey-Vogt, S.~Forrest, and W.~Weimer.
\newblock A systematic study of automated program repair: Fixing 55 out of 105
  bugs for \$8 each.
\newblock In {\em Software Engineering (ICSE), 2012 34th International
  Conference on}, pages 3--13. IEEE, 2012.

\bibitem{DBLP:conf/icse/GouesDFW12}
C.~Le~Goues, M.~Dewey{-}Vogt, S.~Forrest, and W.~Weimer.
\newblock A systematic study of automated program repair: Fixing 55 out of 105
  bugs for {\textdollar}8 each.
\newblock In {\em Proceedings of the 34th International Conference on Software
  Engineering}, pages 3--13, 2012.

\bibitem{LeGoues15tse}
C.~{Le Goues}, N.~Holtschulte, E.~K. Smith, Y.~Brun, P.~Devanbu, S.~Forrest,
  and W.~Weimer.
\newblock The manybugs and introclass benchmarks for automated repair of c
  programs.
\newblock {\em IEEE Transactions on Software Engineering (TSE), in press},
  2015.

\bibitem{le2012genprog}
C.~Le~Goues, T.~Nguyen, S.~Forrest, and W.~Weimer.
\newblock Genprog: A generic method for automatic software repair.
\newblock {\em Software Engineering, IEEE Transactions on}, 38(1):54--72, 2012.

\bibitem{Long15}
F.~Long and M.~C. Rinard.
\newblock Staged program repair with condition synthesis.
\newblock In {\em Proceedings of ESE/FSE}, 2015.

\bibitem{DBLP:conf/pldi/LongSR14}
F.~Long, S.~Sidiroglou{-}Douskos, and M.~C. Rinard.
\newblock Automatic runtime error repair and containment via recovery
  shepherding.
\newblock In {\em {ACM} {SIGPLAN} Conference on Programming Language Design and
  Implementation, {PLDI} '14, Edinburgh, United Kingdom - June 09 - 11, 2014},
  page~26, 2014.

\bibitem{lu2005bugbench}
S.~Lu, Z.~Li, F.~Qin, L.~Tan, P.~Zhou, and Y.~Zhou.
\newblock Bugbench: Benchmarks for evaluating bug detection tools.
\newblock In {\em Workshop on the Evaluation of Software Defect Detection
  Tools}, 2005.

\bibitem{astor}
M.~Martinez and M.~Monperrus.
\newblock {ASTOR: Evolutionary Automatic Software Repair for Java}.
\newblock Technical report, {Inria}, 2014.

\bibitem{DBLP:journals/ese/MartinezM15}
M.~Martinez and M.~Monperrus.
\newblock Mining software repair models for reasoning on the search space of
  automated program fixing.
\newblock {\em Empirical Software Engineering}, 20(1):176--205, 2015.

\bibitem{DBLP:conf/icse/MartinezWM14}
M.~Martinez, W.~Weimer, and M.~Monperrus.
\newblock Do the fix ingredients already exist? an empirical inquiry into the
  redundancy assumptions of program repair approaches.
\newblock In {\em Proceedings of the 36th International Conference on Software
  Engineering}, pages 492--495, 2014.

\bibitem{mechtaev2015directfix}
S.~Mechtaev, J.~Yi, and A.~Roychoudhury.
\newblock Directfix: Looking for simple program repairs.
\newblock In {\em Proceedings of the 37th International Conference on Software
  Engineering}. IEEE, 2015.

\bibitem{monperrus2014critical}
M.~Monperrus.
\newblock A critical review of automatic patch generation learned from
  human-written patches: essay on the problem statement and the evaluation of
  automatic software repair.
\newblock In {\em Proceedings of the 36th International Conference on Software
  Engineering}, pages 234--242. ACM, 2014.

\bibitem{nguyen2013semfix}
H.~D.~T. Nguyen, D.~Qi, A.~Roychoudhury, and S.~Chandra.
\newblock Semfix: Program repair via semantic analysis.
\newblock In {\em Proceedings of the 2013 International Conference on Software
  Engineering}, pages 772--781. IEEE Press, 2013.

\bibitem{noor2015test}
T.~Noor and H.~Hemmati.
\newblock Test case analytics: Mining test case traces to improve risk-driven
  testing.
\newblock In {\em Proceedings of the IEEE 1st International Workshop on
  Software Analytics}, pages 13--16. IEEE, 2015.

\bibitem{DBLP:journals/tse/0001FNWMZ14}
Y.~Pei, C.~A. Furia, M.~Nordio, Y.~Wei, B.~Meyer, and A.~Zeller.
\newblock Automated fixing of programs with contracts.
\newblock {\em {IEEE} Trans. Software Eng.}, 40(5):427--449, 2014.

\bibitem{DBLP:conf/icsm/QiML13}
Y.~Qi, X.~Mao, and Y.~Lei.
\newblock Efficient automated program repair through fault-recorded testing
  prioritization.
\newblock In {\em 2013 {IEEE} International Conference on Software
  Maintenance}, pages 180--189, 2013.

\bibitem{qi2014strength}
Y.~Qi, X.~Mao, Y.~Lei, Z.~Dai, and C.~Wang.
\newblock The strength of random search on automated program repair.
\newblock In {\em Proceedings of the 36th International Conference on Software
  Engineering}, pages 254--265. ACM, 2014.

\bibitem{qi2015efficient}
Z.~Qi, F.~Long, S.~Achour, and M.~Rinard.
\newblock An analysis of patch plausibility and correctness for
  generate-and-validate patch generation systems.
\newblock In {\em Proceedings of ISSTA}. ACM, 2015.

\bibitem{DBLP:conf/icse/SamirniSAMTH12}
H.~Samimi, M.~Sch{\"{a}}fer, S.~Artzi, T.~D. Millstein, F.~Tip, and L.~J.
  Hendren.
\newblock Automated repair of {HTML} generation errors in {PHP} applications
  using string constraint solving.
\newblock In {\em Proceedings of the 34th International Conference on Software
  Engineering}, pages 277--287, 2012.

\bibitem{Shamshiri2015}
S.~Shamshiri, R.~Just, J.~M. Rojas, G.~Fraser, P.~McMinn, and A.~Arcuri.
\newblock Do automatically generated unit tests find real faults? an empirical
  study of effectiveness and challenges.
\newblock In {\em Proceedings of the International Conference on Automated
  Software Engineering}, 2015.

\bibitem{Smith15fse}
E.~K. Smith, E.~Barr, C.~{Le Goues}, and Y.~Brun.
\newblock Is the cure worse than the disease? overfitting in automated program
  repair.
\newblock In {\em Proceedings of the 10th Joint Meeting of the European
  Software Engineering Conference and ACM SIGSOFT Symposium on the Foundations
  of Software Engineering (ESEC/FSE)}, Bergamo, Italy, September 2015.

\bibitem{relifix}
S.~H. Tan and A.~Roychoudhury.
\newblock relifix: Automated repair of software regressions.
\newblock In {\em Proceedings of ICSE}, 2015.

\bibitem{TaoKKX14}
Y.~Tao, J.~Kim, S.~Kim, and C.~Xu.
\newblock Automatically generated patches as debugging aids: a human study.
\newblock In {\em Proceedings of the 22nd {ACM} {SIGSOFT} International
  Symposium on Foundations of Software Engineering}, pages 64--74, 2014.

\bibitem{nopoljournal}
J.~Xuan, M.~Matias, F.~DeMarco, L.~Sebastian, D.~Thomas, D.~Le~Berre, and
  M.~Monperrus.
\newblock Nopol: Automatic repair of conditional statement bugs in large-scale
  object-oriented programs.
\newblock Under 2nd review at IEEE Transactions on Software Engineering.

\bibitem{xuan2014test}
J.~Xuan and M.~Monperrus.
\newblock Test case purification for improving fault localization.
\newblock In {\em Proceedings of the 22nd ACM SIGSOFT International Symposium
  on Foundations of Software Engineering (FSE)}. ACM, 2014.

\bibitem{zhong2015an}
H.~Zhong and Z.~Su.
\newblock An empirical study on real bug fixes.
\newblock In {\em Proceedings of the 37th International Conference on Software
  Engineering (ICSE)}. IEEE, 2015.

\end{thebibliography}

\end{document}